# Growth control of highly textured Bi$_2$Te$_3$ thin films by pulsed laser deposition

*Damian Brzozowski[1], Yu Liu[1], Karola Neeleman[1], Magnus Nord[2], Ingrid G. Hallsteinsen[1]\**

[1]Department of Materials Science and Engineering, Norwegian University of Science and Technology, Trondheim, Norway

[2]Department of Physics, Norwegian University of Science and Technology, Trondheim, Norway

Two-dimensional materials have attracted growing interest due to their unique electronic properties and potential applications in spintronics. Interfacing strongly spin-orbit-coupled chalcogenides with functional oxides such as perovskites has a particularly high potential. In this work, highly textured Bi$_2$Te$_3$ thin films were deposited on (111) oriented SrTiO$_3$ by pulsed laser deposition. We show that, by careful selection of the temperature and pressure of growth, the film's stoichiometry can be manipulated between direct stoichiometry transfer from the target and tellurium-deficient phases. Optimized pulsed laser deposition enables the growth of films with coalesced, faceted grains with grain sizes reaching up to 430 nm, while preserving crystalline quality comparable to that of molecular-beam-epitaxy-grown films. We show striking differences arising from tuning the laser's pulsing frequency and fluence, which lead to changes in surface roughness, the film's porosity, and grain boundaries, as well as grain shape. Analysis of cross-sectional transmission electron microscopy images reveals a sharp substrate-film interface without atomic intermixing and without the formation of amorphous intermediate layers. The results demonstrate that pulsed laser deposition is a viable method for producing high-quality Bi$_2$Te$_3$ thin films and opens the door to the integration of chalcogenides with perovskites with this growth technique.

**Introduction**

Two-dimensional (2D) materials have been an intriguing topic in materials science research due to their exciting properties arising from the spatial confinement of electrons in one dimension. The first example was graphene with its zero-band gap semimetal character, among its exceptional properties,[1] but over the past few years, the library of 2D-materials has expanded to a wide range of materials. $Bi_2Te_3$ has been widely used in thermoelectric applications due to its high thermoelectric figure of merit ZT at room temperature[21] and has gained attention as a topological insulator (TI).[33] Due to the strong spin-orbit coupling in $Bi_2Te_3$, the conduction band minimum, mainly derived from Bi 6p orbitals, and the valence band maximum, derived from Te 5p orbitals, cross each other, leading to band inversion near the high symmetry Γ point. For a topological insulator, the bulk states are insulating, while the surface states remain conducting and are protected by time-reversal symmetry. By introducing magnetic dopants or interfacing $Bi_2Te_3$ with magnetic layers, which break time-reversal symmetry, the quantum anomalous Hall effect is observed at around 2 K,[4] and recently has been observed up to 10 K in exfoliated and mechanically stacked layers of $BiSbTeSe_2$ and magnetic $Cr_2Ge_2Te_6$.[55]

To further take advantage of the intriguing physics of $Bi_2Te_3$ based devices, precise growth control is critical. Fabrication methods such as chemical vapor deposition,[66] electrochemical deposition,[77] molecular beam epitaxy (MBE),[88] sputtering,[99] pulsed laser deposition (PLD),[1010] exfoliation[1111] and others[121] have been extensively utilized. The largest obstacle that remains for the fabrication of 2D material-based devices is the need for a controlled fabrication method that yields continuous smooth thin films of high crystalline quality. Most of the work on chalcogenide growth for topological insulators involves MBE due to its ability to achieve a high crystalline quality within grains, while PLD has been primarily employed for thermoelectric applications, which require thicker layers (over 1 µm) and hence higher growth rates.[1010,12,13] The advantage of PLD lies in its relative simplicity and flexibility as films can be deposited directly from stoichiometric targets. PLD is a cost- and time-efficient technique and is widely used in industrial processes.[1414] Critically, for the deposition of chalcogenides, PLD is argued to offer better crystalline quality at lower deposition temperatures than sputtering techniques because it guarantees higher diffusivity and a greater number of atoms in excited states during plasma formation.[1515]

A remaining challenge in all growth methods is the defect density, and disorder at the interface. For integration with magnetic materials, disorder at the interface decreases the magnetic proximity effect, while magnetic doping leads to non-uniform energy gaps, both of which lead to very low working temperatures for magnetic topological insulators.[16][15] In addition, defects in the topological insulator material, such as grain boundaries and stacking faults, provide parasitic conduction paths masking the topological currents.[17] Because of the high vapor pressure of tellurium and its volatility, $Bi_2Te_3$ films are prone to tellurium deficiency, resulting in the formation of undesirable BiTe and mixed phases.[13,18,19] These non-stoichiometric phases render the bulk p-type or n-type semiconducting. Hence, ensuring insulating bulk states is a key challenge for the utility of these materials. This study aims to understand the influence of various PLD growth parameters in overcoming these challenges, resulting in films of high crystalline quality with lower defect density.

A recently used material as a substrate for chalcogenides is strontium titanate, $SrTiO_3$.[20][20] The crystal structure of $SrTiO_3$ is cubic, with the space group Pm-3m (221) (Figure 1(a), (b)) with lattice parameter $a = 3.905$ Å. The (111) crystallographic plane (Figure 1(c)) exposes six-fold crystal surface symmetry, the same as that of $Bi_2Te_3$ along its (001) direction (Figure 1(d)). Bismuth telluride, $Bi_2Te_3$, crystallizes in the rhombohedral, tetradymite-type crystal structure, presented in Figure 1(e), (f), with the space group R-3m (166), and lattice parameters $a = 4.386$ Å, $c = 30.497$ Å, and $\gamma = 120°$.[21][21] Its unit cell consists of three quintuple layers (QLs) stacked along the $c$-axis, with the atomic configuration $Te^1$-Bi-$Te^2$-Bi-$Te^1$ within a QL. Given the distance between neighboring titanium atoms along (111)-$SrTiO_3$ equal 5.522 Å, the lattice mismatch between (001)-$Bi_2Te_3$ and (111)-$SrTiO_3$ is 20.47%. However, there is a coincidental lattice match upon multiplying the $Bi_2Te_3$ $a$-lattice parameter five-fold, and the $SrTiO_3$ Ti-Ti distance four-fold, which equals 0.59%. Coincidental lattice sites have been shown to give orientational growth of BTs grains and could potentially be used to control nucleation sites and decrease defects.[22] $SrTiO_3$ is also highly insulating which is beneficial for transport measurements of TIs. Until now, most of perovskite-grown chalcogenides were deposited using MBE.[23][23-26][26] Growth by sputtering has also been demonstrated.[27][27] To the authors' best knowledge, no growth of $Bi_2Te_3$ thin films on perovskite substrates by PLD has been reported so far.

In this work, we present PLD growth control of $Bi_2Te_3$ thin films on (111)-$SrTiO_3$. Considering the high volatility of bismuth telluride's elements and its sensitivity to high temperatures, we make use of the main advantages of PLD and show that it is a suitable choice for the high-quality, low-temperature growth of chalcogenides. Growth changes arising from manipulating deposition temperature, pressure, laser fluence, and laser frequency are investigated, where we show striking differences arising from tuning the laser's pulsing frequency and fluence which lead to changes in surface roughness, the film's porosity, and grain boundaries, as well as grain shape. In general, slowing down the kinetics of the growth is paramount to achieving smooth, high-crystalline, stoichiometric films. Finally, optimized growth conditions provide $Bi_2Te_3$ films with large, faceted grains of more than 400 nm, uniform substrate coverage, precise stoichiometry control, and crystalline quality comparable to high-quality MBE-grown films. Investigation of the interface confirms precise control over the formation of defects and the stacking mode of the $Bi_2Te_3$ layers and adding a Te-seed layer results in a sharp interface without intermixing or amorphous layers.

**Experimental**

$Bi_2Te_3$ (BT) thin films were deposited onto (111)-$SrTiO_3$ (STO) substrates by PLD. The substrates (SurfaceNet) were first sonicated in acetone, ethanol, and deionized water at room temperature for 5 min each, followed by etching in a buffered hydrofluoric acid solution (1:9 ratio of HF and $NH_4F$) for 1 minute to ensure single termination. Next, they were annealed at 1050°C for 1 hour in an oxygen flow, resulting in a step-and-terrace surface structure.[2828] Within the PLD chamber the substrates were further annealed at 280°C in an oxygen pressure of 0.3 mbar for 1 h to eliminate organic contaminants, and ensure a pristine surface. A KrF excimer laser, with wavelength $\lambda = 248$ nm, was used to ablate a stoichiometric $Bi_2Te_3$ target (Kurt J. Lesker, 99.999%). The target-substrate distance was 50 mm. PLD chamber remained at a background pressure of $2*10^{-8}$ mbar before introducing argon gas with a working pressure varying between 0.1, 0.3, and 1.0 mbar. The substrate temperature was measured with an external pyrometer and set at 220, 270, and 320°C. To investigate the influence of deposition growth rate and adatom kinetic energy on the film quality, laser repetition rate was varied between 0.2 and 10 Hz, and laser fluence between 0.5 and 2 $Jcm^{-2}$. After deposition, BT films were cooled naturally to room temperature at the deposition pressure. For the *Interface optimization* section, a tellurium seed layer was deposited to passivate the substrate, and a tellurium capping layer was deposited onto the BT film to avoid oxygen diffusion to the interface from the atmosphere. Tellurium layers were deposited at room temperature, $2*10^{-7}$ mbar pressure, 1 Hz pulsing frequency, and 1.5 $Jcm^{-2}$ laser fluence.

Structural analysis was carried out by x-ray diffraction (XRD) in Bruker's D8 Discover diffractometer with copper source (Kα $\lambda$ = 1.5406 Å) and Gobel mirror optic in Bragg-Brentano mode. A referential diffraction pattern of a clean (111)-STO substrate is provided in supporting information, Figure 1S. Surface analysis was done by atomic force microscopy (AFM) with Bruker's Dimension Icon and Thermo Fisher Scientific's Apreo scanning electron microscopy (SEM). AFM images were taken with ScanAsyst Air probes with a nominal tip radius of 2 nm. The images are presented in supporting information, Figure 2S. Gwyddion software was used to correct image leveling and evaluate the average grain size and surface root mean square (RMS) values from acquired AFM images. Crystalline size values were estimated with Gwyddion by extracting profiles along arbitrary lines of selected grains and averaged. SEM images were taken in secondary electron mode at 2 kV and 0.2 nA. Raman spectra were taken with Oxford Instruments' alpha300R in grating 1800 g/mm. To prevent BT surface damaging, laser power was set to 1 mW. The data was normalized to the background. The Transmission Electron Microscopy (TEM) lamella was prepared using a Thermo Fischer FEI Helious G4 UX Focused Ion Beam (FIB) using the lift-out method. The sample was sputter coated with 10 nm of conductive Pt/Pd alloy prior to processing in the FIB. The high-resolution TEM data was acquired on a JEOL ARM200CF at 200 kV, in STO's (11-2) zone axis. The Scanning TEM – Electron Energy Loss Spectroscopy (EELS) data was acquired on the same instrument at 200 kV, on a GIF Quantum ER.

**Results and discussion**

*Temperature and pressure*

During PLD growth, the chamber pressure thermalizes adatoms, while the substrate temperature determines the mobility of adatoms at the surface. To compare the changes in $Bi_2Te_3$ thin films under various pressure and substrate temperature conditions, SEM images are presented in Figure 2. At the highest temperature, 320°C (Figure 2(c), (f), (i)) there is a high degree of desorption. The deposited material does not form continuous films, and instead it forms large, isolated particles. At 0.1 mbar the particles are spherical, while for 0.3 mbar they resemble plates with hexagonal features, the corner angles equal 120°, and at 1.0 mbar we observe hexagonal structures of up to 1 µm in diameter. At 220 and 270°C for all pressures, continuous films are formed, and they consist of a mixture of flat grains and grains that are tilted with respect to the substrate normal, indicating partial misalignment along the *c*-axis. The grains do not show any clear in-plane preferential orientation.

The films grown at the lowest pressure 0.1 mbar, 220 and 270°C (Figure 2(a), (b)) exhibit voids between BT grains; hence, the films have a certain porosity. The 0.1 mbar – 220°C film (Figure 2(a)) shows a particularly high degree of tilted BT grains. Increasing pressure to 0.3 mbar (Figure 2(d), (e)) and further on to 1.0 mbar (Figure 2(g), (h)) leads to more uniform substrate coverage and agglomeration of grains.

The average grain size and RMS acquired from AFM images (Figure S2 of supporting information) are compared and presented in Table 1 (left side). The grain size of the films increases from 100 to 320 nm with the temperature and pressure. Comparably, the change in the RMS value across the series is smaller. RMS ranges between 14.6 to 24.6 nm among the low- and intermediate-temperature samples and increases at higher pressures. The density of tilted grains goes down at higher pressures, but they are taller, contributing more to the higher RMS.

Crystallinity was studied with x-ray diffraction. Figure 3 (left side) shows XRD patterns collected at *2θ* from 15 to 60 degrees in the Bragg-Brentano geometry. All films are highly textured, consisting mainly of the (00l) peaks, and are crystalline which is confirmed by sharp diffraction peaks. Desirably, the dominant peaks correspond to the (003n) crystallographic plane family, confirming that the preferential plane orientation is along BT's crystallographic *c*-axis. Several samples reveal a low intensity peak at 27.64°, which, for the 2:3 stoichiometrical ratio, corresponds to the (105) plane. The (105) plane, along with the (003n) planes, is thermodynamically the most preferred orientation due to its relatively low surface free energy and can be suppressed under optimal thin film growth conditions.[21]

Stoichiometry is analyzed using Raman spectroscopy (Figure 3, right side). For $Bi_2Te_3$, the main points of interest are the peaks located at the shifts of 103 and 133 $cm^{-1}$. The shifts indicate $E^2_g$ and $A^2_{1g}$ vibrational modes, where the $E^2_g$ is the in-plane optical phonon mode, whereas $A^2_{1g}$ is the atom vibration out-of-plane with respect to the quintuple layer base.[29,30][29][30] The $A^2_{1g}$ shift at 133 $cm^{-1}$ is characteristic to the $Bi_2Te_3$ phase (2:3 elemental ratio). The changes in the Raman spectra, including the presence of additional peaks can point to a given stoichiometry or co-existence of phases.[31][31,32] [32] As shown by Concepción et al., Raman spectroscopy proves to be a simple yet powerful tool for qualitative analysis of the stoichiometry of Bi-Te phases.[33][33]

The Raman spectra of the 320°C, 0.3 and 1.0 mbar sample, shown in Figure 3(b), exhibit shifts located at 92, and 120 cm$^{-1}$, as well as a weak signal at 103 cm$^{-1}$, strongly indicating Te deficiency and the presence of the BiTe phase. Although the 92 cm$^{-1}$ peak cannot be identified with certainty, it may belong to the Bi A$_{1g}$ mode.[32][32] On the XRD spectra, Figure 3(a), the 0.3 and 1.0 mbar films show a low-angle peak located at 18.49°, which corresponds to the BiTe phase. Accordingly, the peaks at 27.59°, 45.32°, and 56.86° of the 0.3 mbar sample correspond to the (104), (0 0 12), and (0 0 15) BiTe peaks, respectively. The signal from the 320°C, 0.1 mbar sample (yellow line) is too weak to observe any bismuth telluride features. Considering the elevated substrate temperature of 320°C, it is most likely that the volatile Te species re-evaporate from the STO surface, leading to Bi-rich phases. Hence the isolated hexagonal particles seen on SEM images are mainly BiTe.

For the samples grown at 270°C, we observe a change in stoichiometry in Raman spectra (Figure 3(d)) with respect to pressure. The high- and intermediate-pressure samples (purple and blue lines) reveal the Bi$_2$Te$_3$ Raman peaks at 103 and 133 cm$^{-1}$, while the 0.1 mbar pressure sample spectrum (yellow line) reveals a significant intensity decrease of the E$^2_g$ mode and disappearance of the A$^2_{1g}$ mode. Moreover, peaks at 92 and 124 cm$^{-1}$ are also present. They correspond to the A$^1_{1u}$ and A$^2_{1u}$ vibrational modes, respectively, and represent the crystal-symmetry breaking vibrations along the c-axis (out-of-plane), indicating a BiTe phase. The spectrum of this sample suggests a mixed phase of Bi$_2$Te$_3$ and BiTe. In the XRD diffractograms of the 270°C films (Figure 3(c)), high-intensity (003n) peaks of the Bi$_2$Te$_3$ phase are present. Additionally, secondary peaks are observed for the 1.0 and 0.1 mbar samples. The peak of the 1.0 mbar sample located at 27.67° corresponds to the (105) crystallographic orientation. The 0.1 mbar sample shows a peak at 16.73°, marked with the red circle, which is discussed further in the next section, *Pulsing frequency and laser fluence*. The decrease of the deposition pressure leads to a relatively larger Te plume propagation than the Bi plume. As the Te plume expands over a larger area, the relative Te concentration at the substrate decreases with respect to the Bi species, promoting the growth of the BiTe phase.[13] Indeed, as compared with the study by Concepción et al., the Raman spectra suggest mixed phase formation.[33][33] Although SEM images for 270°C show good control over the substrate coverage, the film at low pressure suffers from tellurium losses and mixed stoichiometry.

Figure 3(f) shows the Raman spectra for the 220°C samples. The spectra provide strong evidence for the 2:3 stoichiometry, with shifts present only at 103 and 133 cm$^{-1}$. A noticeable decrease in the A$^2_{1g}$ peak intensity relative to $E^2_g$ is observed as the deposition pressure increases. It is possible that increasing deposition pressure leads to thicker films due to enhanced interaction of the vapor phase with the chamber gas, confining the ablated material to a narrower plume. As the film undergoes the transition from thin-film-like to bulk-like, the out-of-plane lattice vibrations become increasingly more constrained, as suggested by Shahil et al., leading to reduced $A^2_{1g}$ intensity.[34] XRD of the 220°C films, Figure 3(e), reveals the (003n) planes of the $Bi_2Te_3$ phase. A (105) peak is present for the 0.1 mbar sample, and an unidentified peak for the 0.3 mbar sample of low intensity. The peak does not correspond to the BiTe phase; however, it may indicate the presence of a Bi-Te phase with a reduced *c*-axis lattice parameter, possibly due to structural defects.

The (006) peaks of $Bi_2Te_3$ (and (005) peaks of BiTe) were used to estimate the *c*-axis lattice parameter and FWHM. The data is summarized in Table 2 (top section). All measured c-axis lattice parameters are smaller than the reference value of 30.497 Å. Similarly, for BiTe, the measured *c* values are lower than the theoretical $c_{BiTe}$ = 24.050 Å, possibly indicating Te deficiency. For *2θ*, there is a clear trend across all pressures, indicating lattice compression with increasing deposition temperature. At the same time, for a constant temperature, there is a slight decrease in *2θ* with decreasing pressure. No clear trend is observed for FWHM, which ranges between 0.61° and 0.19°. The lowest FWHM value, and the *c* value closest to the reference, belong to the low-temperature, high-pressure sample, which also shows the best stoichiometry as confirmed by Raman spectroscopy.

Based on the above-discussed XRD and Raman spectra, it is evident that tuning the deposition pressure affects not only the crystallinity of bismuth telluride phases but also their stoichiometry. This observation is especially crucial when deposition is performed without a tellurium-rich source or flow control, as is typically implemented in MBE. Similarly, the growth temperature influences the sticking coefficient of adatoms, increasing the probability of desorption at elevated temperatures. To obtain highly crystalline film with proper stoichiometry, the deposition temperature must be tuned to a range in which tellurium loss from the substrate is negligible, while still allowing for sufficient surface mobility of adatoms for reorganization. In our case, this temperature window is found at 220°C, which is lower than the 250-350°C range reported for BT PLD growth on various substrates.[13,18,35] A similar temperature window of 200-250°C was found for growth on graphite.[36] Temperatures reaching 400°C have been shown to yield optimal films when an additional tellurium source is introduced.[37] To display a more insightful view into stoichiometry and structural quality of BT thin films, electrical transport measurements can be taken as a complementary analysis to the Raman spectroscopy. Electrical transport properties greatly rely on the density of structural defects. In the case of $Bi_2Te_3$, these defects take a non-linear dependence on growth conditions, i.e., substrate temperature. While the density of n-type tellurium vacancies $V_{Te}$ steadily increases with the substrate temperature, the opposing p-type $Bi_{Te}$ antisite defects have the lowest formation energy and become dominant at elevated substrate temperatures under tellurium-deficient conditions.[48,49] Changes to the type of conductivity, as well as other transport properties, in the function of temperature, would provide a more detailed perspective onto the structural properties and their influence on functional properties.

The analysis further shows that decreasing temperature is critical for preserving the 2:3 elemental ratio in the PLD deposition, while increasing the pressure up to 1.0 mbar promotes high quality crystalline growth. Here, we observe the (006) peak FWHM value of 0.19° at 1.0 mbar, 220°C sample, comparable to the reports on high-quality BT films grown with various techniques, where FWHM ranges between 0.12-0.60°.[18,35,38,391839] Although pressures above 1.0 mbar were not explored in this work, it is reasonable to expect that further pressure increases eventually lead to a more disoriented assembly of the BT crystallites and potentially inhomogeneous stoichiometric transfer of Bi and Te species.[13]

*Pulsing frequency and laser fluence*

In addition to the analysis of temperature and pressure, the influence of laser parameters on film growth was investigated. A series of four samples was grown at a temperature of 220°C, and a pressure of 1.0 mbar, based on the results from the above-discussed sample series. The laser fluence was set to 0.5 and 1.5 $Jcm^{-2}$, whereas the laser frequency was set to 0.2 and 10 Hz. Qualitative stoichiometry analysis with Raman spectra indicates 2:3 elemental ratio for all samples (see Figure S3 of supporting information).

Figure 4 shows SEM images of the grown films. Similarly, as for the 220°C, 1.0 mbar sample discussed in the previous section, there is a very high degree of substrate coverage with little to no voids observed. However, distinct changes are evident in the grain arrangement and shape. The high frequency samples (Figure 4(a), (b)) reveal a rougher surface topography than the low frequency samples (Figure 4(c), (d)). The degree of crystallite misorientation out-of-plane is greatly reduced at the low laser frequency. The effect of the laser fluence is observed in the shape of BT grains, where at the low fluence (Figure 4(a), (c)) the edges are sharper and more well-defined. The grains also appear larger and flatter relative to the surface normal. In contrast, the high fluence grains (Figure 4(b), (d)) take on irregular shapes; the films are more porous, with clear boundaries between grains that, unlike in the low fluence samples, are not conglomerated. An important highlight is the presence of spiral-like growth of BT columns in the 0.2 Hz, 0.5 Jcm$^{-2}$ sample, Figure 4(c). An AFM image of the sample was used to take step height profiles (Figure 4(e)). The spiral-shaped columns have a step height of approximately 1.0 nm, corresponding to the height of a BT quintuple layer. Several explanations have been proposed for spiral formation in chalcogenide films, including the presence of screw dislocations and pinning of domains followed by coalescence.[40] Considering that spirals indicate layer-by-layer growth in physical vapor deposition of chalcogenides, it is most likely that the origin of this effect is growth front pinning, which becomes the dominating growth mechanism at slower deposition rates and promotes 2D expansion of grains. This mechanism is also observed in high-quality BT films regardless of substrate choice.[41-44] Grain size and RMS values were measured based on AFM scans (see Figure S4 of supporting information) and summarized in Table 1 (right side). Compared to the 220°C, 1.0 mbar sample from the previous section, the discussed samples yield lower RMS values, ranging from 8.3 to 17.3 nm. The grain sizes remain in the order of hundreds of nanometers. A striking change in grain size is observed for the 0.2 Hz, 0.5 Jcm$^{-2}$ sample, where the estimated size is roughly tripled and exceeds 430 nm. Simultaneously, the same sample has the lowest RMS value. One can conclude that low frequency and low fluence conditions lead to the smoothest surface topography and promote 2D-like film growth.

Wide scan XRD diffractograms are presented in Figure 5(a), and the (006) peak parameters are summarized in Table 2 (bottom section). XRD patterns in Figure 5(a) show that regardless of the laser parameters, all films display high texturing and preferential orientation of growth. For the high-frequency films, peak intensities (yellow and red lines) are considerably lower, and there is a strong contribution of the (105) crystallographic planes as well.

The 0.2 Hz, 1.5 Jcm$^{-2}$ sample (purple line) reveals the (006) peak at 17.53°, with a FWHM of 0.19°. It is noteworthy that the secondary peak appears on the low-angle side of the (006) peak at 17.23°. Tellurium-deficient phases (BiTe, Bi$_4$Te$_3$) do not have peaks at this $2\theta$ value; however, the peak could arise from the presence of a phase with dominant Bi$_{Te}$ antisite defects, which expand the lattice along its $c$-axis.[45] The 0.2 Hz, 0.5 Jcm$^{-2}$ sample (blue line) yields exceptionally high intensity peaks without the presence of secondary phases or non-(003n) peaks. The FWHM of the (006) peak is 0.17°, which is lower than that for any other sample discussed and lower than other PLD-grown BT films on various substrates, e.g., $Al_2O_3$, Si, mica.[18,18,35,38 38] In order to measure film's thickness, x-ray reflectivity (XRR) scan was taken for the 0.2 Hz, 0.5 Jcm$^{-2}$ sample (Figure S5 of supporting information). XRR fringes confirm the high quality of the layer stacking. The fit model estimated the film's thickness to be about 46.4 nm.

To detect phases that are not aligned with respect to the surface normal, grazing incidence XRD (GIXRD) was employed. Figure 5(b) presents GIXRD data of the frequency-fluence sample series. The 10 Hz, 1.5 Jcm$^{-2}$ sample (yellow line) reveals a weak signal from (006) and (105) crystallographic planes. It is the only sample yielding a non-(003n) peak in the GIXRD diffractogram. The 10 Hz, 0.5 Jcm$^{-2}$ (red line) and 0.2 Hz, 1.5 Jcm$^{-2}$ (purple line) diffractograms show exclusively (003n) peaks, indicating that there is a certain number of (003n)-oriented grains that are misaligned with respect to the substrate surface plane. The 0.2 Hz, 0.5 Jcm$^{-2}$ sample (blue line) is the only one not showing peaks on its GI diffractogram. This indicates that BT grains are highly oriented out-of-plane, and that for this film, the degree of unfaceted grains is the lowest and below the diffractogram's detectability.

Considering results of the whole series, laser fluence as low as 0.5 Jcm$^{-2}$ is sufficient for the proper transfer of the 2:3 bismuth telluride phase. In contrast, high laser fluence significantly alters the film morphology. As high fluence results in greater kinetic energy of ablates species, their interaction with previously deposited material becomes much more intense, disrupting the controlled 2D propagation of grains. The result is a dominating Volmer-Weber-type island growth mode, characterized by bulkier, more spherical grains with less-defined shapes. The effect of laser frequency tuning primarily influences the smoothness and surface uniformity of the film. At low frequencies, more time is available for adatoms to diffuse to energetically favorable sites before being disrupted by material from subsequent laser pulses. For bismuth telluride, this effect is clearly observed even at a relatively low substrate temperature of 220°C. In contrast, at high frequencies, adatoms have limited time to diffuse and organize, resulting in a less uniform, rougher coating across the substrate.

*Interface optimization*

Cross-section analysis was performed to investigate the interface quality. A $Bi_2Te_3$ film was grown on (111)-oriented STO, following the optimized conditions for flat, uniform coverage (220°C, 1.0 mbar, 0.2 Hz, 0.5 Jcm$^{-2}$). Two additional steps were introduced: 1) a tellurium seed layer was deposited to passivate the substrate surface;[46,46] 2) a protective tellurium capping layer was deposited onto the $Bi_2Te_3$ film to prevent oxygen diffusion from the atmosphere to the interface.[46,46,47,47] Prior to analyzing the interface of the Te-capped film, a bare BT film was deposited with the Te seed layer to assess the influence of the seed layer on film topography. The film was grown using half the number of laser pulses to evaluate the influence of the seed layer on the initial stages of growth. The SEM image is presented in Figure 6(a). The film consists predominantly of faceted grains approximately 150 nm in size. We observe voids between grains as well as a number of unfaceted grains at the surface. However, because the film was grown with fewer pulses, it represents an early growth stage, and the grain size and porosity should not be directly compared to other samples. Most importantly, the Te seed layer does not negatively impact the film's topography.

A high-resolution cross-sectional TEM image of the Te-capped $Bi_2Te_3$ film, shown in Figure 6(b), reveals laterally stacked, uniform BT layers, each approximately 1 nm thick, corresponding to the QL thickness. The interface region, shown in Figure 6(c) and marked with an arrow, indicates that the QLs are parallel to the substrate surface, and no amorphous interfacial phase is observed. Fast Fourier transform (FFT) analysis was performed in this region to determine the epitaxial relationship between the substrate and the film, Figure 6(d). The diffraction spots correspond to the characteristic planes of STO and $Bi_2Te_3$, and no secondary phases are detected. This confirms that $Bi_2Te_3$ (003n) planes grow parallel to the (111) surface plane of STO, demonstrating the highly ordered nature of $Bi_2Te_3$ layers. Figure 6(e) shows EELS line profiles acquired across the interface. Within experimental uncertainty, no clear atomic intermixing is observed between the film and the substrate. The interface region is estimated to be approximately 5 nm thick. Although slight intensity changes are seen at 19 and 22 nm for all elements, they remain within the measurement uncertainty due to various experimental factors. Furthermore, the overlap of Te and O edges in the EELS spectra complicates the elemental analysis in the given resolution. Within the resolution of our measurement techniques the TEM images show a sharp interface between the two layers starting with a Te-seed layer.

*Outlook*

The main challenge with $Bi_2Te_3$ as a TI towards electronic applications is that the bulk is not insulating enough, but rather exhibits a bulk semiconducting behavior which overshadows the surface current. In general, this is believed to be due to inhomogeneities in the material which introduces excess carriers. It has been shown that a decrease in point defects leads to higher electrical surface conductivity and mobility.[50] The dominant $V_{Te}$ point defects in $Bi_2Te_3$ are a source of charge carriers whose high concentration leads to increased carrier scattering, effectively lowering mobility. As noted, a measured n-type/p-type conduction will tell if the majority of point defects are tellurium vacancies $V_{Te}$ or $Bi_{Te}$ antisite point defects.[48,49] The relationship between films' microstructure and electrical transport is on the other hand less understood, and only a few reports are presented on the role of individual inhomogeneities on the parasitic bulk conduction in $Bi_2Te_3$. Kwang-Chon et al.[51] argue that twin domains suppress mobility by creating extra charge carriers. The influence of grain sizes is less clear, with some reports suggesting little influence over transport properties,[52] while others show decreased resistivity with larger grains.[50]

At room temperature, MBE grown samples show electrical mobility between 10-100 $cm^2V^{-1}s^{-1}$.[2] At temperatures below 50 K, a review summary done by Ngabonziza et al. displays samples with electrical mobility between 700-1900 $cm^2V^{-1}s^{-1}$ for $SrTiO_3$-grown films, and 35-5800 $cm^2V^{-1}s^{-1}$ across other substrates.[52] The huge differences in transport between reports come not only from the nature of point defects, but films' thickness, surface oxidation and contact material [ref] as well. Samples with structural characteristics similar to those presented here—namely high crystallinity, controlled stoichiometry, and moderate grain sizes—typically exhibit room-temperature mobilities of 35–50 $cm^2V^{-1}s^{-1}$, however, the thickness reported here is not optimized for transport measurement of the surface currents.

**Conclusions**

In summary, we investigated the influence of key PLD parameters on quality of $Bi_2Te_3$ thin films - substrate temperature, ambient pressure, laser fluence, and laser frequency. In general, slowing down the kinetics of the growth is paramount to achieving smooth, high-crystalline, stoichiometric films. We show that by carefully controlling the substrate temperature, high-quality, stoichiometric $Bi_2Te_3$ films can be grown without the need for a Te-rich source. It is achieved through control of adatoms' kinetic energy, reducing desorption probability.

Moreover, high deposition pressures (1.0 mbar) stabilize the stoichiometric transfer of high vapor pressure species such as Te and Bi. Tuning these parameters affects film porosity and grain size, yielding grains in the 100-320 nm range. Upon investigating the effect of laser parameters, we observe that film topography can be significantly altered while preserving target stoichiometry. Lowering the pulsing frequency to 0.2 Hz reduces surface roughness by allowing adatoms more time to diffuse to energetically favorable sites. In contrast, increasing the frequency to 10 Hz results in non-uniform coatings with a greater presence of unfaceted grains. Reducing the laser fluence leads to compact films with coalesced grains, while high fluence produces porous structures with droplet-like grain shapes and clearly defined grain boundaries. Notably, the grain size is only marginally affected, and high crystallinity can be preserved. Most importantly, by combining low frequency and low fluence, we achieved highly faceted $Bi_2Te_3$ films with an exceptional grain size exceeding 430 nm, along with distinct spiral growth features, consistent with the van der Waals-layered nature of $Bi_2Te_3$. Cross-section HRTEM show that with a Te-seed layer we achieve a sharp interface between the two materials, with no evidence of intermixing or amorphous layers. We demonstrate that PLD is a powerful and tunable deposition method for growing $Bi_2Te_3$ thin films with precise control over morphology, stoichiometry, and crystallinity. Under optimal conditions, PLD enables fabrication of films with low surface roughness, high crystalline quality, large grain size and sharp interfaces. This study provides valuable insights into the growth control of other chalcogenide topological insulators (i.e., $Bi_2Se_3$, $MnBi_2Te_4$) via PLD and paves the way for fabricating $Bi_2Te_3$-based heterostructures on other perovskite materials with functional properties.

FIGURES

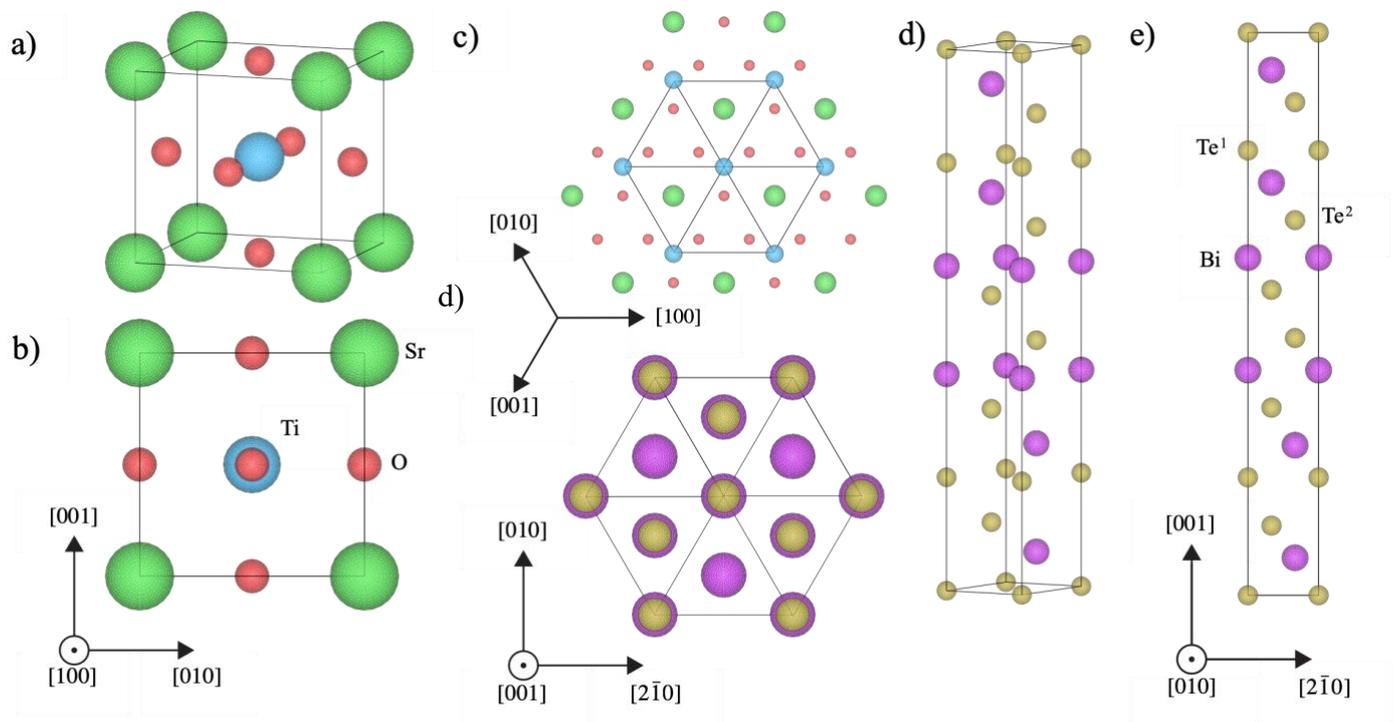

Figure 1. a) Bi$_2$Te$_3$ unit cell. b) Bi$_2$Te$_3$ side view with labelled elements. c) SrTiO$_3$ unit cell. d) SrTiO$_3$ side view with labelled elements. e) Projection of SrTiO$_3$ (top) and Bi$_2$Te$_3$ (bottom) showing six-fold symmetry.

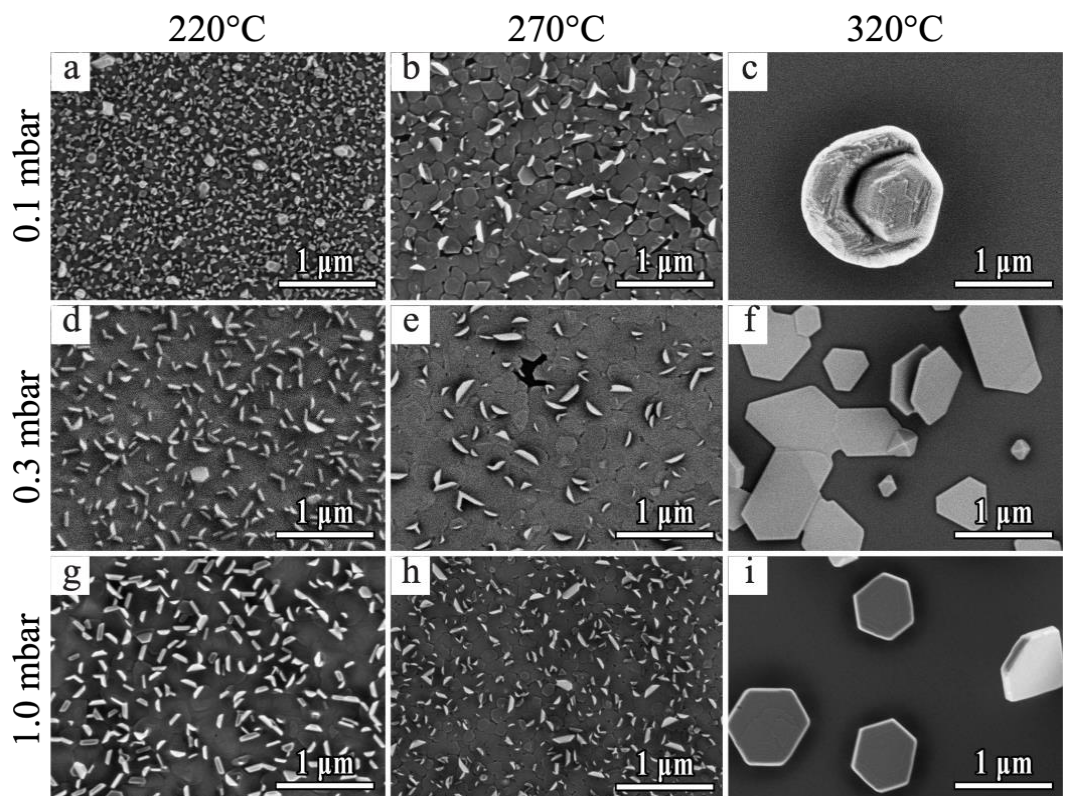

Figure 2. SEM images of SrTiO$_3$-Bi$_2$Te$_3$ samples of the temperature-pressure series. The images were taken in the secondary electron mode.

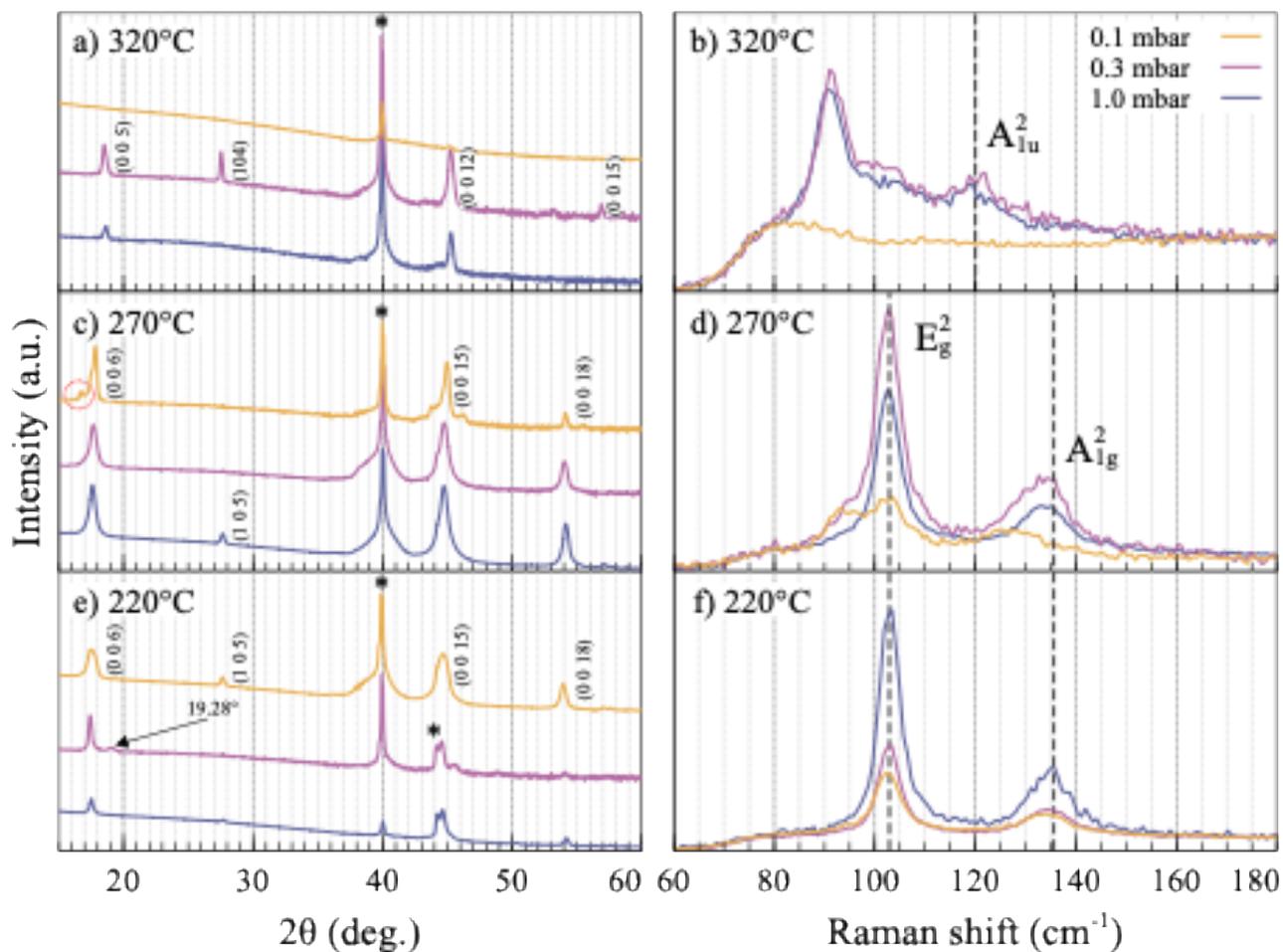

Figure 3. Left side: XRD $\theta$-$2\theta$ scans of Bi$_2$Te$_3$ films, plotted in logarithmic scale. Substrate peaks are marked with asterisk. Right side: Raman spectra of the temperature-pressure sample series. Main vibration modes are labelled. The scans are sorted by the deposition temperature.

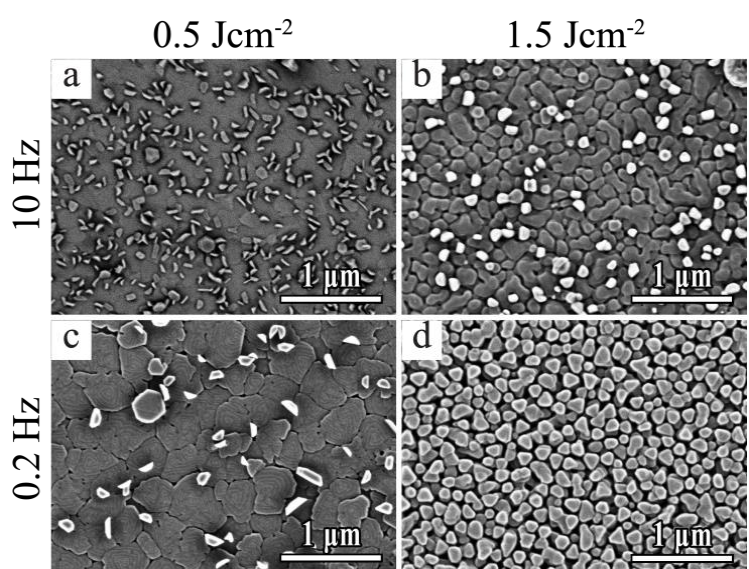

Figure 4. SEM images of the Bi$_2$Te$_3$ films of the frequency-fluence series, deposited at 220°C, 1.0 mbar Ar. Images taken at the secondary electron mode.

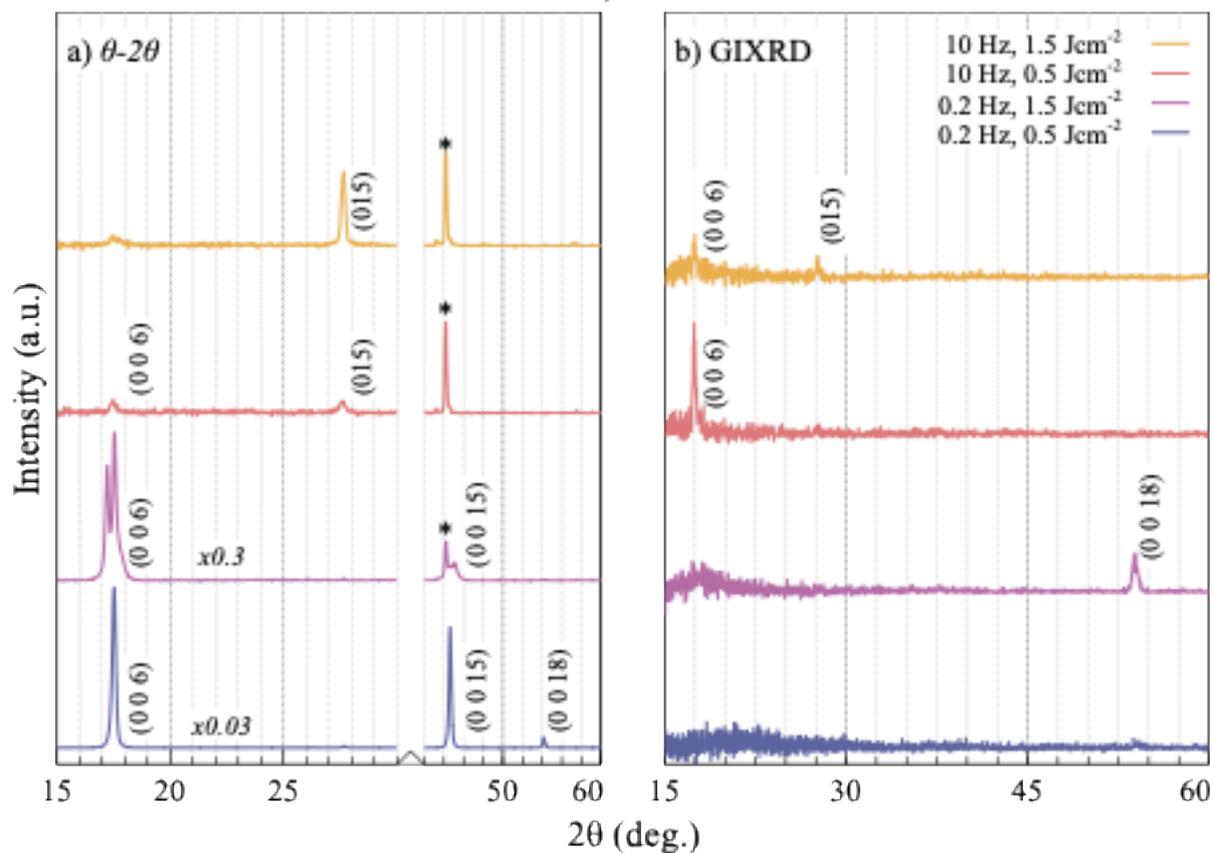

Figure 5. Left side: XRD θ-2θ scans of $Bi_2Te_3$ films grown at 220°C, 1.0 mbar Ar. The range corresponding to the main STO peak is cropped. A secondary substrate peak at 44.2° is marked with asterisk. Diffraction data of low frequency samples is downscaled for readability. Right side: Grazing incidence XRD of $Bi_2Te_3$ films of the frequency-fluence series, grown at 220°C, 1.0 mbar Ar.

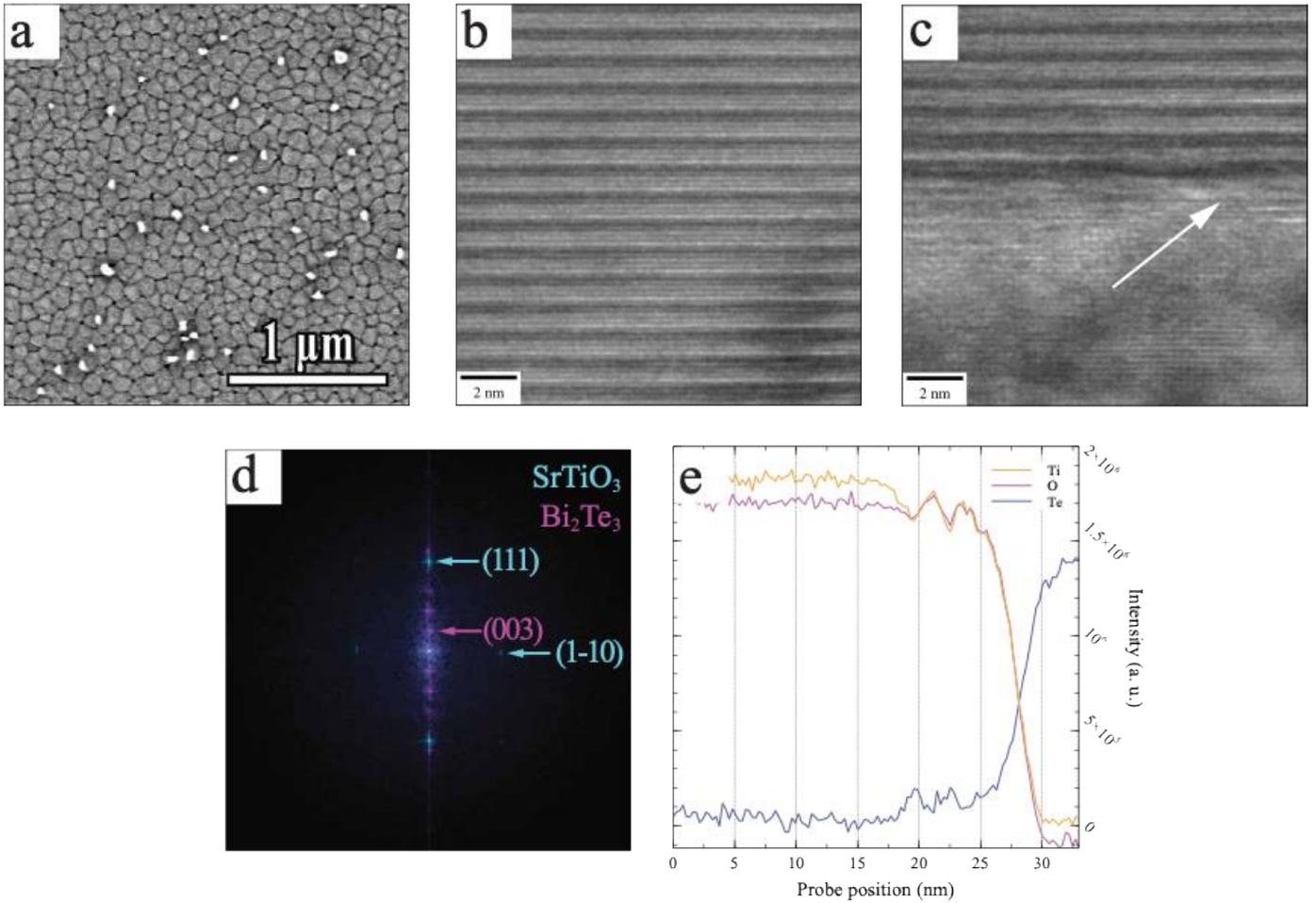

Figure 6. a) SEM image of BT thin ~~capped~~ film with Te seed layer. BT film deposited at 220°C, 1.0 mbar, 0.2 Hz, 0.5 Jcm$^{-2}$. b) High-resolution cross-sectional TEM of Bi$_2$Te$_3$ film grown on STO, with zone axis (100) of Bi$_2$Te$_3$. c) HRTEM of the STO-BT interface with zone axis (11-2) of STO. The arrow marks the interface. d) FFT of the interfacial region. The peaks are color-coded and assigned to identified phases. e) EELS profiles of titanium, oxygen, and tellurium across the interface.

TABLES.

Table 1. Grain size and root mean square values estimated from AFM scans of the temperature-pressure and frequency-fluence series

| | Grain size, RMS (nm) | | | | | |
|---|---|---|---|---|---|---|
| | *220°C* | *270°C* | *320°C* | | *0.5 Jcm$^{-2}$* | *1.5 Jcm$^{-2}$* |
| *0.1 mbar* | 103, 17.1 | 184, 14.6 | -- | *10 Hz* | 135.6, 17.3 | 188.4, 14.4 |
| *0.3 mbar* | 157, 16.8 | 247, 21.1 | 695, 52.6 | *0.2 Hz* | 430.2, 8.3 | 161.4, 11.4 |
| *1.0 mbar* | 202, 20.5 | 320, 24.6 | -- | | | |

Table 2. XRD data collected for the temperature-pressure and frequency-fluence series, taken from the (006) BT peak. The c-axis lattice parameter is calculated based on the 2θ position of the peak. Parameters marked with asterisk refer to the c-axis of BiTe phase and correspond to its (005) peak

| | | **XRD (006)$_{BT}$** | | |
|---|---|---|---|---|
| *Temperature (°C)* | *Pressure (mbar)* | *2θ (deg.)* | *$c_{BT}$ (Å)* | *FWHM (deg.)* |
| 320 | 0.1 | -- | -- | -- |
| 320 | 0.3 | 18.58 | 23.86* | 0.32 |
| 320 | 1.0 | 18.68 | 23.73* | 0.29 |
| 270 | 0.1 | 17.86 | 29.77 | 0.20 |
| 270 | 0.3 | 17.72 | 30.01 | 0.35 |
| 270 | 1.0 | 17.61 | 30.19 | 0.29 |
| 220 | 0.1 | 17.66 | 30.11 | 0.61 |
| 220 | 0.3 | 17.54 | 30.31 | 0.21 |
| 220 | 1.0 | 17.53 | 30.33 | 0.19 |
| *Frequency (Hz)* | *Fluence (Jcm$^{-2}$)* | *2θ (deg.)* | *$c_{BT}$ (Å)* | *FWHM (deg.)* |
| 10 | 1.5 | 17.49 | 30.40 | -- |
| 10 | 0.5 | 17.49 | 30.40 | -- |
| 0.2 | 1.5 | 17.54 | 30.31 | 0.19 |
| 0.2 | 0.5 | 17.53 | 30.33 | 0.17 |

ASSOCIATED CONTENT

The following files are available free of charge:

Brzozowski2025 supporting information (PDF) containing:

- Substrate's XRD
- AFM scans of samples, used for estimations in Table 1
- Raman spectra of the second series of samples
- XRR of the low fluence-low frequency sample used for the thickness estimation

AUTHOR INFORMATION

**Corresponding Author**

*Ingrid G. Hallsteinsen – Department of Materials Science and Engineering, Norwegian University of Science and Technology, Trondheim, Norway; https://orcid.org/0000-0003-0789-8741; Email: ingrid.g.hallsteinsen@ntnu.no

**Author Contributions**

D.B.: Methodology, formal analysis, investigation, data curation, and writing-original draft. Y.L.: Resources. K.N.: Resources. M.N.: Resources. I.G.H.: Project administration, funding acquisition, conceptualization, writing-review, editing, and supervision. ‡These authors contributed equally. (match statement to author names with a symbol)

**Funding Sources**

This project was supported by the Norwegian Research Council under Project Number 325063.

**Notes**

The authors declare no competing financial interest.

ACKNOWLEDGMENT

This project was supported by the Research Council of Norway under Project Number 325063. The Research Council of Norway is acknowledged for the support to the Norwegian Micro- and Nano-Fabrication Facility, NorFab, project number 295864. The authors acknowledge the support from the Research Council of Norway for NORTEM (197405) and InCoMa (315475).

ABBREVIATIONS

2D, two-dimensional; AFM, atomic force microscopy; BT, bismuth telluride; EELS, electron energy loss spectroscopy; FFT, fast Fourier transform; FIB, focused ion beam; FWHM, full-width half maximum; GIXRD, grazing incidence x-ray diffraction; MBE, molecular beam epitaxy; PLD, pulsed laser deposition; QL, quintuple layer; RMS, root mean square; SEM, scanning electron microscopy; STO, strontium titanate; TEM, transmission electron microscopy; TI, topological insulator; XRD, x-ray diffraction.

*For Table of Contents Use Only*

# Growth control of highly textured $Bi_2Te_3$ thin films by pulsed laser deposition

*Damian Brzozowski[1], Yu Liu[1], Karola Neeleman[1], Magnus Nord[2], Ingrid G. Hallsteinsen[1]\**

[1]Department of Materials Science and Engineering, Norwegian University of Science and Technology, Trondheim, Norway

[2]Department of Physics, Norwegian University of Science and Technology, Trondheim, Norway

SYNOPSIS

We demonstrate precise pulsed laser deposition (PLD) control for high-quality $Bi_2Te_3$ thin films on (111)-$SrTiO_3$ substrates. By optimizing temperature, pressure, laser fluence, and frequency, we achieved stoichiometric, smooth, and highly crystalline films with large grains and sharp, defect-free interfaces. This work establishes PLD as a tunable route for chalcogenide topological insulator growth.

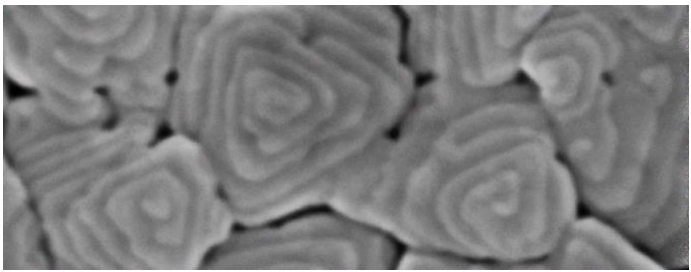

# Supporting Information

# Growth control of highly textured $Bi_2Te_3$ thin films by pulsed laser deposition

*Damian Brzozowski[1], Yu Liu[1], Karola Neeleman[1], Magnus Nord[2], Ingrid G. Hallsteinsen[1]\**

[1]Department of Materials Science and Engineering, Norwegian University of Science and Technology, Trondheim, Norway

[2]Department of Physics, Norwegian University of Science and Technology, Trondheim, Norway

*ingrid.g.hallsteinsen@ntnu.no

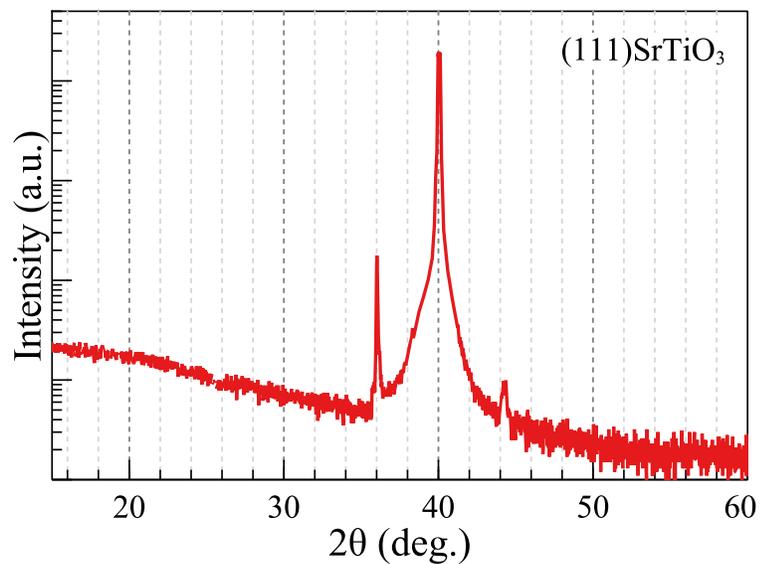

**Figure S1.** XRD of (111)-SrTiO$_3$ substrate, plotted in logarithmic scale. The substrate was subjected to pre-deposition treatment, which involved hydrofluoric acid etching and high-temperature annealing.

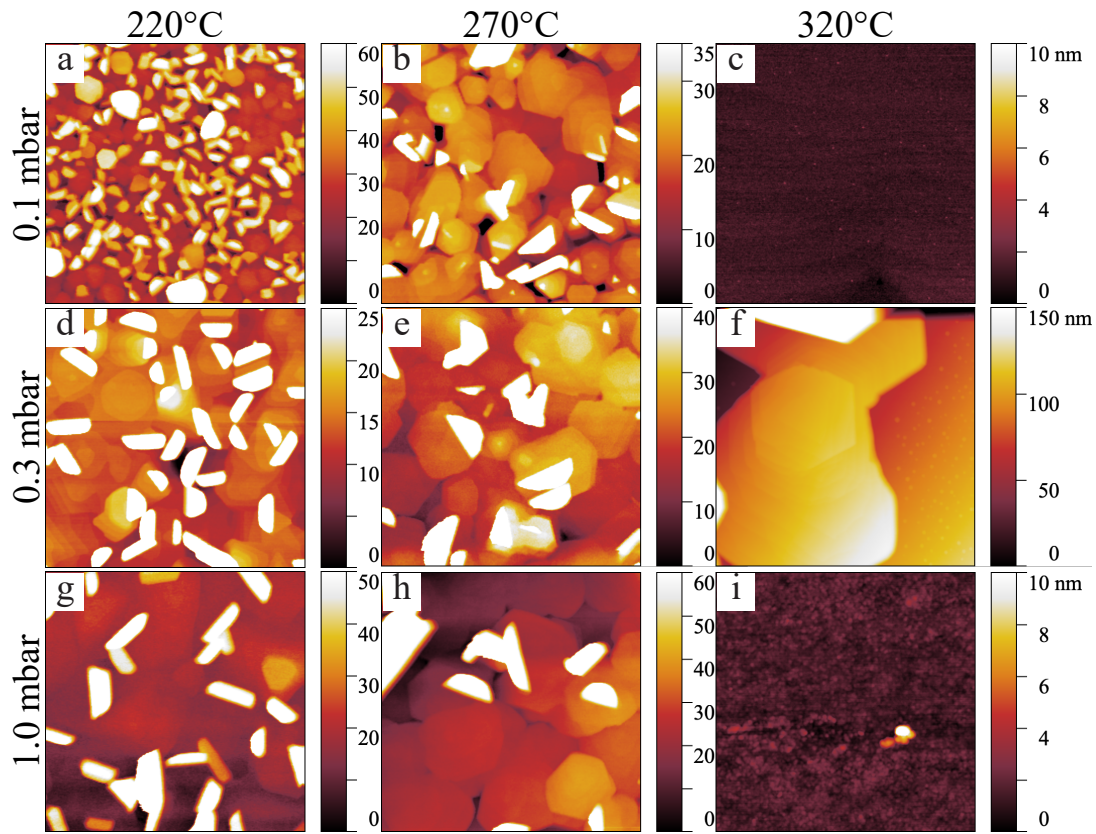

**Figure S2.** AFM images of SrTiO$_3$-Bi$_2$Te$_3$ samples, 1 × 1 μm grown at 220°C/270°C/320°C and 0.1/0.3/1.0 mbar. For each sample, the same number of laser pulses was sent, and the laser parameters remained constant.

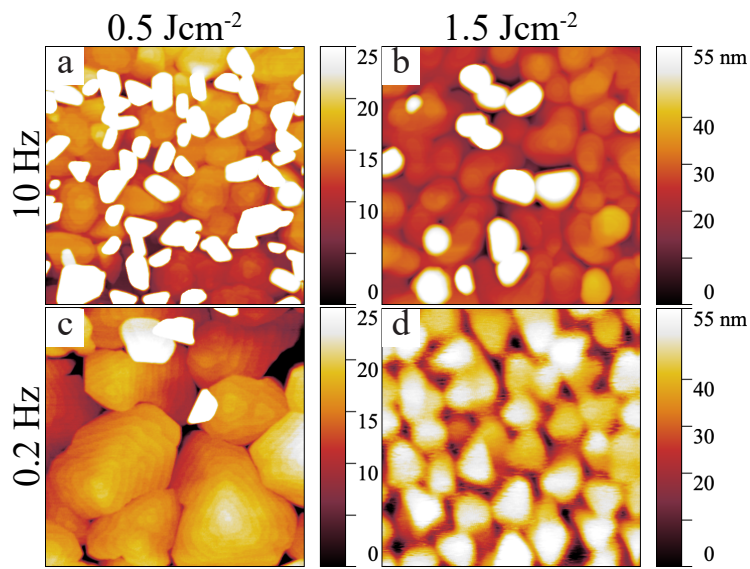

**Figure S3**. AFM images of $Bi_2Te_3$ films of the frequency-fluence series, deposited at 220°C, 1.0 mbar Ar.

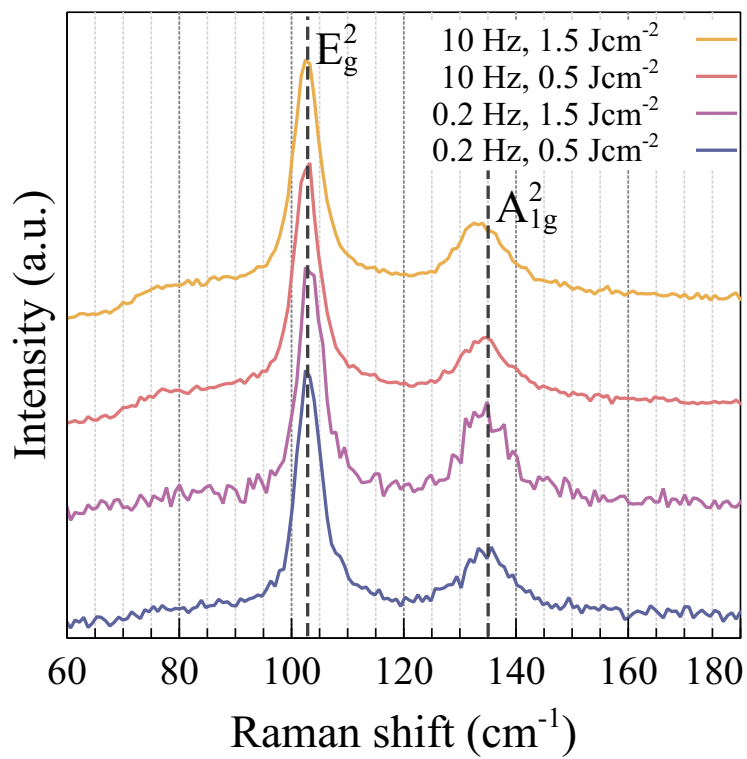

**Figure S4.** Raman spectra of the frequency-fluence sample series. The spectra were normalized to the $E^2_g$ peak.

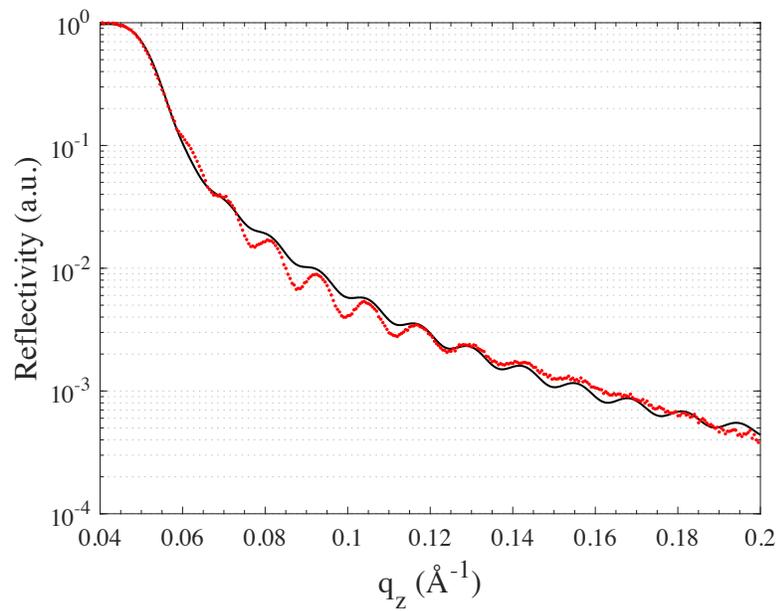

**Figure S5**. X-ray reflectivity profile of the low fluence-low frequency sample. Measurement data is plotted with red dots; the fit used for thickness estimation is plotted with the black line.